\documentclass{article}
\usepackage{chicago}
\newcommand{\vctr}[1]{\ensuremath{\mathbf{ #1 }}}

\newcommand{\dr}[1]{\ensuremath{\mathrm{d} #1\,}}
\newcommand{\mc}[1]{\ensuremath{\mathcal{#1}}}

\newcommand{\pbp}[2]{\ensuremath{\frac{\partial #1}{\partial #2}}}

\newcommand{\op}[1]{\ensuremath{\widehat{\textsf{\ensuremath{#1}}}}}

\newcommand{\be}{\begin{equation}}
\newcommand{\ee}{\end{equation}}
\newcommand{\e}[1]{\mathrm{e}^{#1}}

\newcommand{\iec}{\mbox{i.\,e.\,}}

\usepackage{graphicx}

\begin{document}

\title{Deflating the Aharonov-Bohm Effect}
\author{David Wallace\thanks{Balliol College, Oxford; email: david.wallace@balliol.ox.ac.uk}}
\maketitle

\begin{abstract}
I argue that the metaphysical import of the Aharonov-Bohm effect has been overstated: correctly understood, it does not require either rejection of gauge invariance or any novel form of nonlocality. The conclusion that it does require one or the other follows from a failure to keep track, in the analysis, of the complex scalar field to which the magnetic vector potential is coupled. Once this is recognised, the way is clear to a local account of the ontology of electrodynamics (or at least, to an account no more nonlocal than quantum theory in general requires); I sketch a possible such account.
\end{abstract}

\section{Introduction}

In classical electromagnetism, the magnetic field can be represented either by the field strength $\vctr{B}$, or by a vector field \vctr{A} such that $\nabla \times \vctr{A}=\vctr{B}$, where in the latter case $\vctr{A}$ is determined only up to a family of transformations known as \emph{gauge transformations}. Prior to the discovery --- and empirical confirmation --- of the Aharonov-Bohm (A-B) effect, it was possible to believe (and, I think, widely \emph{was} believed) that \vctr{A} had only mathematical significance and that a true description of the magnetic field required only $\vctr{B}$. The A-B effect demonstrated  --- as uncontroversially as anything in the foundations of physics --- that there are features of electromagnetism that transcend the local action of the magnetic field strength on charged matter: electrons can move through a region of space in which $\vctr{B}=0$ but which surrounds a region of nonzero $\vctr{B}$, and their behaviour is dependent upon the value of $\vctr{B}$ in that latter region. Mathematically speaking these results are possible because the quantum mechanics of electromagnetism involves the interaction of a complex field $\psi$ with the \vctr{A}-field, and the equations that govern that interaction --- though gauge-independent --- cannot be rewritten in a local way via $\vctr{B}$ alone.

But just what the conceptual import is remains controversial. In foundational discussions of late it has been argued --- and widely accepted --- that the effect requires either that we accept some new form of non-locality beyond that already encountered in quantum mechanics, or that we abandon the principle that gauge transformations simply redescribe the same physical goings on. In particular, the A-B effect rests on the fact that the values of $\vctr{B}$ within a spatial region need not determine the field $\vctr{A}$ in that region even up to gauge transformations --- but that the residual gauge-invariant features of $\vctr{A}$ not captured by $\vctr{B}$ have an inherently local character to them.  

In this paper I argue that much of this debate\footnote{Including some parts to which I contributed: cf \citeN{wallacetimpsonshort}.} rests upon a mistake: that of considering the $\vctr{A}$-field in isolation rather than in conjunction with the $\psi$-field. After reviewing the A-B effect and the contemporary foundational literature in section~\ref{review}, I demonstrate this in section~\ref{ABcomplex} by considering the gauge-invariant features of $\psi$ and \vctr{A} jointly, which are not exhausted by the gauge-invariant features of $\psi$ and $\vctr{A}$ separately. I demonstrate that those joint features can in general be given an entirely local characterisation, blocking the concern that some gauge-invariant features are inherently non-local. In section~\ref{origins} I show in a different way how this apparent nonlocality arises in the study of \vctr{A} alone and how it is blocked when we allow for \vctr{A} and $\psi$ jointly.

In section \ref{interlinking} I attempt an interpretation of these results: my proposal is that we should not think of $\psi$ and $\vctr{A}$ as representing separate entities but as representing, jointly and redundantly, features of a single entity, with the redundancy being localisable either to $\psi$ or to $\vctr{A}$ as a matter of pure convention; I illustrate this proposal via brief consideration of the Higgs mechanism. 

After a brief discussion of the generalisation of my analysis to non-Abelian gauge theories in section \ref{nonabelian} (this section could be skipped on a first reading), in sections \ref{vestiges}-\ref{interactingwithwhat} I address two possible concerns with the account I give, and in doing so explore further the extent to which we can give a properly local account of the physical goings on around the solenoid in the A-B effect. Section~\ref{conclusion} is the conclusion.

\section{The A-B Effect Reviewed}\label{review}

The classical theory of a point electric charge moving under the influence of a background magnetic field is straightforward. The particle is represented mathematically by a vector function $\vctr{q}(t)$ of time, and the field by a vector field $\vctr{B}(\vctr{x},t)$. The field satisfies two of Maxwell's equations,
\be \label{maxwell}
\nabla \cdot \vctr{B}(\vctr{x},t) =0 \,\,\,\,\,\mbox{and}\,\,\,\,\nabla \times \vctr{B}(\vctr{x},t)= 4 \pi \vctr{J}(\vctr{x},t),
\ee
where $\vctr{J}$ is the electric current density,
 and the force on it is given by the Lorentz force law,
\be 
\vctr{F}(t) = e \dot{\vctr{q}}(t)\times \vctr{B}(\vctr{q},t),
\ee
where $e$ is the particle's charge. (I use Gaussian units with $c=1$.) In general we will be working in the background-field regime, where the back-reaction of the particle on the field is ignored.

Mathematically, it is always possible to express $\vctr{B}$ as the curl of another vector field $\vctr{A}$, the \emph{vector potential}: $\vctr{B} = \nabla \times \vctr{A}$. In many cases in classical magnetostatics, doing so can be mathematically convenient. For instance, since the divergence of a curl is always zero, the first equation in~(\ref{maxwell}) is automatically satisfied if $\vctr{B}$ is defined in terms of $\vctr{A}$. More relevantly for our purposes, the standard way to put the Lorentz force law into Hamiltonian form uses the Hamiltonian
\be 
H(\vctr{q},\vctr{p})=\frac{1}{2m}(\vctr{p}+e \vctr{A}(\vctr{q}))^2.
\ee
That is: it is expressed in terms of the vector potential, rather than the field strength.

At least in classical electromagnetism, the standard assumption is that $\vctr{A}$ is merely a mathematical convenience, and that $\vctr{B}$ fully represents the physical features of the magnetic field. There are two interrelated reasons for this:
\begin{enumerate}
\item The definition of $\vctr{A}$ in terms of $\vctr{B}$ specifies $\vctr{A}$ only up to the gradient of an arbitrary smooth function $\Lambda$: if we replace $\vctr{A}$ with $\vctr{A}'=\vctr{A}+\nabla \Lambda$, then $\nabla \times \vctr{A}'=\nabla \times \vctr{A}$.
\item Only $\vctr{B}$ appears to be physically detectable.
\end{enumerate}
In the Maxwell equations and the Lorentz force law, the dependence of the physics on $\vctr{B}$ alone rather than $\vctr{A}$ is manifest. It is only tacit in the Hamiltonian formulation of the theory (there is no straightforward way to write a Hamiltonian form of the Lorentz law in terms of $\vctr{B}$ alone), but it is strongly suggested by the fact that the \emph{classical gauge transformation}
\be 
\vctr{A}\longrightarrow \vctr{A}+ \nabla \Lambda;\,\,\,\,\, \vctr{q}\rightarrow \vctr{q};\,\,\,\,\,\vctr{p}\rightarrow \vctr{p}- e \nabla \Lambda
\ee
is a symmetry of the Hamiltonian, and furthermore, a symmetry that leaves the trajectory of the particle unchanged.

In applications of the vector potential in electromagnetism, it is common to impose some additional condition --- a \emph{choice of gauge} --- such that exactly one \vctr{A}-field is compatible with any given set of empirical data. A common choice, for instance, is the Coulomb gauge, defined by the conditions that $\nabla \cdot \vctr{A}=0$ and that $\vctr{A}$ vanishes at spatial infinity. If $\vctr{A}$ and $\vctr{A}'$ are two gauge-equivalent vector potentials related by a gauge transformation $\Lambda$ and both satisfying the Coulomb gauge condition, then $\nabla^2\Lambda=0$, which together with the boundary condition entails that $\Lambda$ is constant and hence that $\vctr{A}=\vctr{A}'$.

The quantum mechanics of a particle interacting with a background magnetic field is obtained in the standard way by replacing $\vctr{q}$ and $\vctr{p}$ in the classical Hamiltonian with the quantum-mechanical position and momentum operators. The resultant Schr\"{o}dinger equation (in units where $\hbar=1$) in the position representation is
\be 
\pbp{\psi}{t}(\vctr{x},t)= \frac{i}{2m} \left(\nabla -i e \vctr{A}(\vctr{x},t)\right)^2 \psi(\vctr{x},t).
\ee
The Schr\"{o}dinger equation is invariant under a quantum-mechanical version of the classical gauge transformation. Since momentum in configuration-space wave mechanics is given by the gradient of the phase of the wave-function, we would expect that the classical momentum transformation becomes a phase change, and so it does: the form of the transformation is
\be \label{gt}
\vctr{A}\longrightarrow \vctr{A}+ \nabla \Lambda;\,\,\,\,\, \psi\longrightarrow \e{i e\Lambda}\psi,
\ee
again for an arbitrary smooth function $\Lambda$. And just as the classical transformation left particle trajectories unchanged, the quantum version leaves unchanged the probability of finding the particle in any given location after any given time.

The gauge-invariance of the Schr\"{o}dinger equation might suggest that, in quantum just as in classical mechanics, it is the \vctr{B}-field rather than the $\vctr{A}$-field that is of physical significnance. The Aharonov-Bohm effect calls this into question: in its simplest form, it works as follows.
\begin{enumerate}
\item A beam of charged particles is separated into two; the two beams flow round opposite sides of a solenoid and are then allowed to re-interfere.
\item In the absence of any current through the solenoid (and hence of any induced magnetic field), there will be a set of interference fringes produced by the reinterference of the two beams.
\item When the solenoid is turned on, there will be a shift in the interference pattern. The magnitude of the shift will be proportional to the difference of the integrals of the $\vctr{A}$-field along the paths traversed by the left and right beams respectively. That is, the shift $\Delta$ will be proportional to the integral of $\vctr{A}$ around the loop $\Gamma$ formed by the two halves of the beam:
\be 
\Delta \propto \oint_\Gamma \! \vctr{A}\cdot \mathrm{d}\vctr{x}
\ee
\item By Stokes' theorem, the line integral of a vector field $\vctr{V}$ around a closed loop in a simply-connected region (that is: a region in which any closed loop can be continuously deformed to a point without moving any part of it out of the region) is equal to the surface integral of the curl of $\vctr{V}$ over any surface bounded by the loop. Since $\nabla \times \vctr{A}=\vctr{B}$, this means that $\Delta$ is proportional to the integral of the magnetic field over the interior of the region enclosed by the beam, or in other words that it is proportional to the magnetic flux through that region.\footnote{Of course, the electron will be quite delocalised, and indeed this delocalisation is central to the observation of interference fringes, so ``the'' path taken by the electron is not really well-defined. But since \vctr{B} vanishes outside the solenoid, by Stokes' theorem any two paths which pass the solenoid on the same side will have the same line integral of \vctr{A}.}
\end{enumerate}
The conceptual problem is that a sufficiently well-constructed and well-shielded solenoid will result both in negligible magnetic field \emph{outside} the solenoid, and negligible wavefunction \emph{inside} the solenoid. So the electron is moving (almost) entirely through a region in which the magnetic field is zero --- and yet, its evolution is still detectably different from what would occur if the solenoid were turned off.

If we hold on to the idea that the magnetic field is completely represented by the field strength $\vctr{B}$ (what \citeN[p.54]{healeybook} calls a `no new EM properties' view), this means action at a distance: the passage of the electron around the solenoid is affected by the magnetic flux within the solenoid directly, without any mediating field to transmit its influence. This is doubly embarrassing because the equations governing the electron's motion certainly \emph{look} as if they involve local action --- but between $\psi$ and $\vctr{A}$, not $\psi$ and $\vctr{B}$.\footnote{There is a subtler problem: the problems of interpretation of the vector potential in electromagnetism generalise to so-called `non-Abelian gauge-theories' (cf section \ref{nonabelian}), but the no new properties view does not generalise readily to these more exotic cases. See~\citeN[p.84]{healeybook} and references therein for details.}

This suggests a natural alternative(called the ``new localized EM properties'' view by \citeN[p.55]{healeybook}): take the \vctr{A}-field as a physical field. The problem, of course, is gauge invariance: since two gauge-equivalent $\vctr{A}$-fields (that is, two $\vctr{A}$-fields related by a gauge transformation) are empirically indistinguishable, how is it to be determined which is the true $\vctr{A}$ field?
This can be thought of as giving rise both to a problem of empirical inaccessibility of the present electromagnetic state (no amount of evidence can tell us which of the various gauge-equivalent \vctr{A}-fields is correct) and a problem of indeterminism (the equations of electromagnetism determine a system's evolution only up to gauge transformations, so if $\Lambda(\vctr{x},t)=0$ for $t<0$, they fail to tell us whether a given set of $t<0$ initial conditions will evolve into $\vctr{A}$ or $\vctr{A}+\nabla \Lambda$).

These are not unfamiliar problems in the foundations of physics: expressed at this level of abstraction, they comprise an electromagnetic version of general relativity's ``hole argument'' \cite{earmannortonhole} where the Einstein field equations suffice to determine the evolution of the spacetime metric only up to diffeomorphisms. There, the standard response\footnote{Here I gloss a substantial literature, of course; see \citeN{Norton2008} and references therein for further discussion} has been to regard diffeomorphically related metrics as different descriptions of the same underlying physics. The analogous strategy in electromagnetism would be to take \emph{gauge}-equivalent \vctr{A}-fields as different representations of the same underlying ontology.

This observations suggests looking for an explicitly gauge-invariant representation of that ontology. Our slogan might be: ``the physical facts about the fields are represented by the gauge-invariant features of $\vctr{A}$. One of those gauge-invariant features is $\vctr{B}=\nabla \times \vctr{A}$ , but the A-B effect shows us that there are others.'' As stated, this is a mathematical problem: find a complete characterisation of $\vctr{A}$, up to gauge transformations, in any given region $R$. And there is a well-known answer: $\vctr{A}$ is characterised completely and gauge-invariantly by its line integral around every loop in $R$ (called the \emph{holonomies} of the loops).

For future purposes, it will be useful to explain this a little further. Given some functional $f$ from $\vctr{A}$-fields to some other space, $f$ can be said to characterise the gauge-invariant features of the $\vctr{A}$-fields provided that $f(\vctr{A})=f(\vctr{A}')$ iff $\vctr{A}$ and $\vctr{A}'$ are related by a gauge transformation. To see that this is the case for holonomies, suppose that $\vctr{A}$ and $\vctr{A}'$ satisfy
\be 
\oint_\Gamma \!\vctr{A}\cdot \mathrm{d}\vctr{x}=\oint_\Gamma \!\vctr{A}'\cdot \mathrm{d}\vctr{x}
\ee
for any loop $\Gamma$. Then the integral of $(\vctr{A}-\vctr{A}')$ around any closed loop is zero, or put another way, the integral of $(\vctr{A}-\vctr{A}')$ between $\vctr{x}_0$ and $\vctr{x}$ depends only on $\vctr{x}_0$ and $\vctr{x}$ and not on the path connecting them. If we then choose arbitrary $\vctr{x}_0$ and define
\be 
\Lambda(\vctr{x})=\int^\vctr{x}_{\vctr{x}_0}(\vctr{A}-\vctr{A}')\cdot \mathrm{d}\vctr{x},
\ee
then $\nabla \Lambda=(\vctr{A}-\vctr{A}')$ and so $\vctr{A},\vctr{A}'$ are gauge-equivalent. Conversely, if they are gauge-equivalent then (since the integral of $\nabla \Lambda$ around a closed loop always vanishes) they have the same holonomies.

This suggests Healey's own preferred interpretation of the magnetic field's ontology, the ``new non-localized EM properties'' view: the magnetic field is represented by a map from loops to real numbers. By the definition of the curl, the integral of $\vctr{A}$ around an infinitesimal loop at point $\vctr{x}$ is equal to $\vctr{B}\cdot \vctr{n}\delta S$, where $\vctr{n}$ is normal to the surface enclosed by the loop and $\delta S$ is the area of that surface. So among the components of Healey's ontology (in effect) is the magnetic field. But that ontology is much larger than just the field.

Healey's loop ontology faces three main objections. Firstly, just as with the $\vctr{B}$-field ontology there is no natural way to write the equations of motion of the theory in terms of the loop properties directly; the $\vctr{A}$-field remains indispensable mathematically. Secondly, the ontology is very redundant: loops can be decomposed into smaller loops, and the real number assigned to the larger loop must be the sum of those assigned to its components. (If a region $R$ is simply connected, any loop can be decomposed into infinitesimal loops, and the $\vctr{B}$ field of $R$ actually completely determines the values of all the loops in $R$.) Not only is this awkward, it is difficult to explain naturally \emph{except} by defining the values of each loop as the integral of some vector field around the loop.

Most strikingly, Healey's ontology is non-separable: if $X$ and $Y$ are simply connected spatial regions whose union is not simply connected, then fully specifying the values assigned to each loop in $X$ and $Y$ separately leaves some loops in $X\cup Y$ unspecified. The A-B effect itself offers an illustration: consider $X$ and $Y$ to be as given in diagram 1. Since $X$ and $Y$ are each simply connected, and since in each $\vctr{B}=0$, each is magnetically trivial: each loop integral is equal to zero. Insofar as the magnetic field in a region is supposed to be represented by the gauge-invariant facts about $X$ in that region, in both $X$ and $Y$ the magnetic field is the same as in empty space (there is a gauge transformation that transforms it to zero). But the field in $X\cup Y$ is \emph{not} the same as in empty space: the value of loops that enclose the solenoid is non-zero.

\begin{figure}[h]
\textbf{\caption{Regions of space around the solenoid}}
\includegraphics[width=4in,height=4in, trim=0in 4in 1in 0in, keepaspectratio=true]{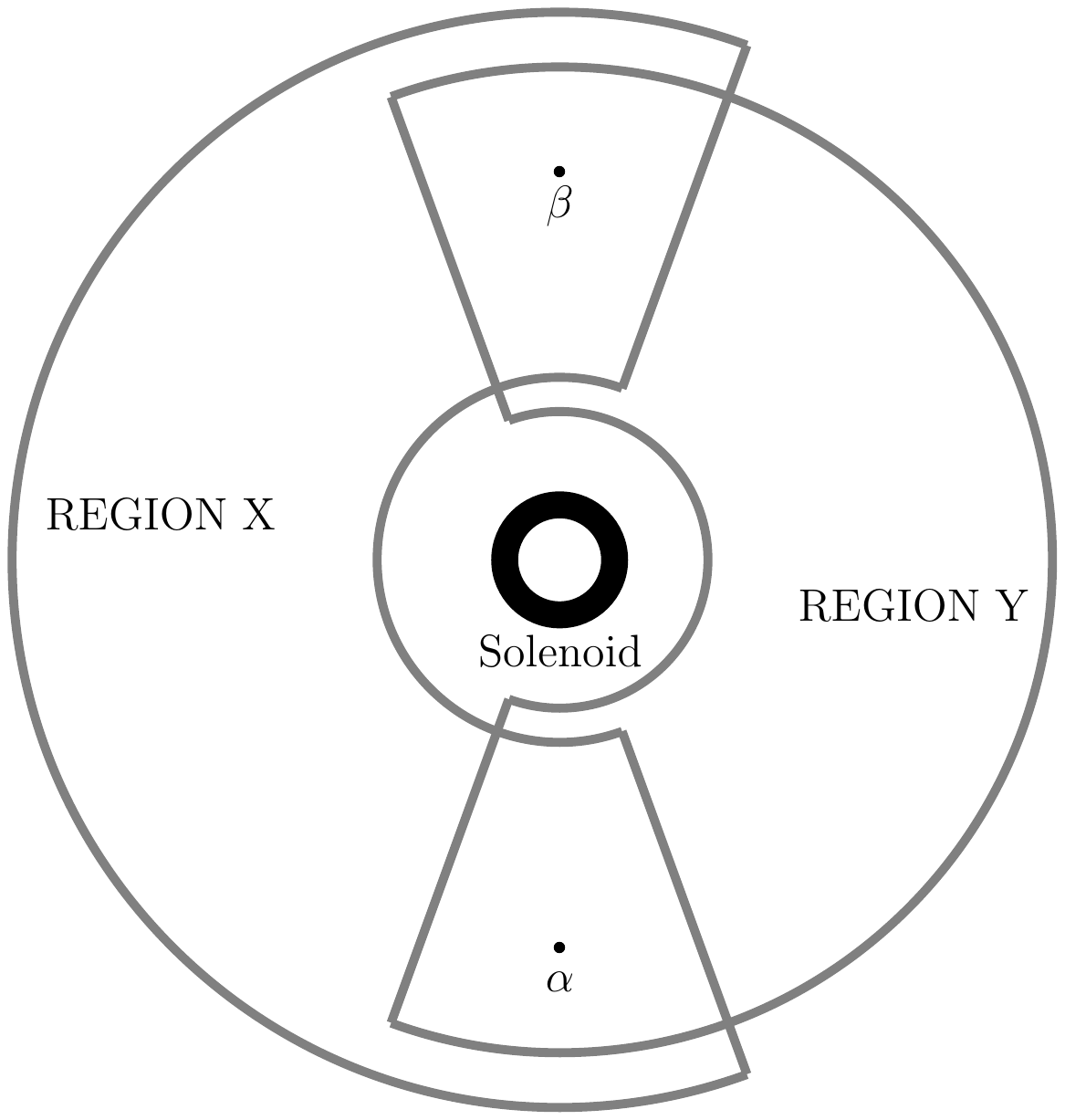}
\end{figure}

Could this simply be an awkward feature of Healey's choice of representation? Other representations of gauge-invariant features certainly exist: indeed, somewhat counter-intuitively, any \emph{choice of gauge} is a gauge-invariant representation of the theory. For consider: a choice of gauge selects, out of each equivalence class of \vctr{A}-fields under gauge transformations, exactly one member of that class, so $\vctr{A}$ and $\vctr{A}+\nabla \Lambda$ give rise to the same field with respect to any given gauge. We could, for instance, represent the gauge-invariant features of the ontology simply by writing the \vctr{A}-field in the Coulomb gauge.

It is, however, readily \emph{provable} that any such representation is non-separable (I give such a proof in section \ref{origins}.)  A corollary is that no choice of gauge can actually be specified locally: knowing \vctr{A} in some proper subregion $X$ of space does not determine $\vctr{A}$ on $X$ with respect to any given gauge. (Note that the Coulomb gauge, in addition to the local condition $\nabla \cdot \vctr{A}=0$, requires a boundary condition at infinity.) So the non-separability of Healey's representation is not a pathology of that representation but an inevitable feature of any such representation.

There is an alternative, more robust, way to make a specific choice of gauge that \emph{does} avoid nonlocality. We can simply stipulate by fiat that our theory is represented by some vector field \vctr{A} satisfying a given gauge condition, and that the value of $\vctr{A}(\vctr{x})$ is to be interpreted as representing not some non-local gauge-invariant feature of the electromagnetic field, but some genuinely local feature of $\vctr{x}$. \citeN{maudlingauge} refers to this strategy as the One True Gauge approach.\footnote{In discussion I have found that Maudlin is often understood as advocating this strategy; my own more minimal reading is that he is simply pointing out that it is possible as part of a case to undermine analogies between the A-B effect and Bell's inequality.}

But given the gauge symmetry of the equations of electromagnetism, there seem to be few grounds beyond aesthetic preference for selecting one gauge rather than another, and the problems of empirical inaccessibility and indeterminism are replaced by a problem of underdetermination of theory by data. One need not be a crude verificationist to find this level of underdetermination unattractive.

Furthermore (and this seems to be underappreciated in the literature) there is a certain vestige of nonlocality present even in the ``one true gauge'' approach. For since $\vctr{A}(\vctr{x})$ is underdetermined by any gauge-invariant properties in a neighborhood of $\vctr{x}$, and since only the gauge-invariant quantities are empirically accessible, the One True Gauge approach is local only at the expense of making some ontological features of a region $X$ inaccessible from within $X$ even in principle.\footnote{There is a certain similarity here to \citeN{deutschhayden}'s account of quantum theory in the Heisenberg picture; there, too, all features of the theory are formally local but many are empirically inaccessible without access to spatially distant regions.} (Recall that there cannot exist a choice of gauge without this property, for any such gauge would \emph{per impossibile} determine a separable representation of the gauge-invariant properties.)

In any case, the A-B effect appears to present us with a trilemma. We would like an understanding of electromagnetism that is separable, gauge invariant, and has no action at a distance. It appears that one of these has to be rejected.

Before going on I should note that while this discussion has been carried out at a relatively elementary level, many proposed ways of understanding the ontology of electromagnetism in the light of the A-B effect are much more sophisticated, and in particular, involve extensive appeal to the mathematics of fibre bundles\footnote{See, for instance,  \citeN{leedsgauge} or \citeN{nounougauge}.}. It is perhaps worth making clear that whatever the virtues of these approaches, they cannot avoid the basic trilemma. For the \vctr{A}-field in region X is gauge-equivalent to what it would be if the solenoid were absent, and so is the \vctr{A}-field in region Y, but the \vctr{A}-field in region $X \cup Y$ is not. So any representation of the field that is gauge-invariant must  violate either separability (by assigning a nontrivial electromagnetic state to region $X \cup Y$) or local action (by assigning a trivial electromagnetic state to the region in which the electron moves).

Here ends my summary of the A-B effect.

\section{The A-B effect and the complex field}\label{ABcomplex}

The A-B effect arises because of certain features of the mathematical theory of a complex scalar field $\psi$ coupled to a real vector field \vctr{A}. It is therefore in hindsight a little odd that the literature on the A-B effect has been almost wholly concerned with the $\vctr{A}$ field and hardly at all with the $\psi$ field. In particular, the line of reasoning that leads to the loop ontology --- and to the argument that any gauge-invariant representation of the magnetic field is non-separable --- is concerned purely with the gauge-invariant features of $\vctr{A}$ and not with $\psi$ at all. Let us attempt to rectify this.

Prima facie, there are two obvious ideas as to how to think about the gauge-invariant features of $\psi$:
\begin{enumerate}
\item Representing the gauge-invariant features of $\vctr{A}$ by loop holonomies already takes care of the gauge freedom. Any two complex fields $\psi,\psi'$ can thus be thought of as representing different physical possibilities. The physical states of the theory are thus represented by a complex field and a set of loop holonomies.
\item Since there is a gauge transformation relating any two fields $\psi,\psi'$ satisfying $|\psi(\vctr{x})|=|\psi'(\vctr{x})|$, the only gauge-invariant feature of $\psi$ is its magnitude. The physical states of the theory are thus represented by a real field $|\psi|$ and a set of loop holonomies.
\end{enumerate}
Neither is satisfactory, for neither provides a complete characterisation of the gauge-invariant features of the theory. To see why, suppose that $(\psi,\vctr{A})$ and $(\psi',\vctr{A}')$ are two possible pairs of fields. A given function $f$ of the fields characterises them completely up to gauge transformations provided that $f(\psi,\vctr{A})=f(\psi',\vctr{A}')$ just if for some $\Lambda$, $\psi'=\e{i\Lambda}\psi$ and $\vctr{A}'=\vctr{A}+\nabla \Lambda$.

For the first suggested characterisation, $f$ takes $\psi$ to itself and $\vctr{A}$ to the loop holonomies. But here the only gauge transformations that leave $\psi$ invariant are those for which $\Lambda(\vctr{x})\neq 0$ only when $\psi(\vctr{x})=0$. So in general this representation is not itself gauge invariant. For the second suggestion, $f$ takes $\psi$ to its magnitude and $\vctr{A}$ to its holonomies, and this clearly \emph{is} gauge invariant. But consider the pairs $(\psi,\vctr{A})$ and $(\e{ie\sigma}\psi,\vctr{A})$, for some arbitrary function $\sigma$. These have the same holonomies and the same $|\psi|$. But they are gauge-equivalent only if, for some $\Lambda$,
\be 
\e{ie\sigma}\psi=\e{ie\Lambda}\psi\,\,\,\,\,\mbox{and}\,\,\,\,\,\vctr{A}=\vctr{A}+\nabla \Lambda.
\ee
This pretty clearly requires (i) $\Lambda$ to be constant (at least on the connected parts of the region of space we are considering) and (ii) $\Lambda(\vctr{x})=\sigma(\vctr{x})+2n\pi/e$ on any connected region where $\psi\neq 0$. In general (that is, for any choice of $\sigma$ which is not constant on any connected region where $\psi\neq 0$) this cannot be satisfied. So the second suggested characterisation erroneously represents gauge-inequivalent pairs of fields as physically equivalent. (And, in case it's not obvious, these gauge-inequivalent fields are definitely physically inequivalent: two pairs of fields which at time $t$ are gauge-inequivalent but agree on the magnitude of the wavefunction and on the holonomies will not in general so agree at later times, and $|\psi|$ is empirically accessible.)

Our two suggestions share a common flaw. They attempt to characterise the gauge-invariant features of the fields by separately representing the gauge-invariant features of $\psi$ and $\vctr{A}$. But the gauge transformations act \emph{jointly} on the two fields, and there are joint features of the pair of fields that are gauge-invariant but do not derive directly from gauge-invariant features of the field considered separately. 

In particular, consider the quantity $|\nabla \psi - ei \vctr{A}\psi|$. This is gauge-invariant --- indeed, the fact that it is gauge invariant is the central heuristic of the gauge principle in particle physics\footnote{Slightly more accurately, the central heuristic is that $(\nabla \psi -ei \vctr{A}\psi)$ transforms under the gauge group in the same way as does $\psi$ itself.} --- but its gauge invariance does not derive from gauge-invariant features of $\psi$ and $\vctr{A}$ separately but rather from the cancellation of terms in the gauge transformations of both.

This suggests that a gauge-invariant characterisation of $(\psi,\vctr{A})$ will need to consider joint features. A helpful way to get at such a characterisation starts by decomposing $\psi$ into its magnitude and phase:
\be 
\psi(\vctr{x},t)=\rho(\vctr{x},t)\exp(i e \theta(\vctr{x},t)).
\ee
(This decomposition is unique, up to an overall constant $2n\pi/e$ in $\theta$, provided that $\psi(\vctr{x},t)$ is everywhere nonzero; I return to the $\psi=0$ case later.)

Clearly, $\rho$ is a gauge-invariant feature of $\psi$ alone, and hence of $(\psi,\vctr{A})$ jointly. More interestingly, consider the gauge-invariant quantity $\psi^*(\nabla - ei \vctr{A})\psi$. Expressed in terms of $\rho$ and $\theta$, it is
\be 
\psi^*(\nabla - i \vctr{A})\psi = \rho\nabla \rho + i e\rho^2 (\nabla \theta - \vctr{A}).
\ee
Since $\rho^2$ and $\rho\nabla \rho$ are gauge-invariant, so is $\mc{D}\theta \equiv \nabla \theta - \vctr{A}$, the gauge-covariant derivative of $\theta$ (something that can also be verified directly). 

So: we now have two gauge-invariant features of the theory: the scalar field $\rho=|\psi|^2$, and the vector field $\mc{D}\theta$. In fact, no others are needed. For suppose that $\psi'=\rho'\e{ie \theta'}$ and $\vctr{A}'$ satisfy 
\be \rho'=\rho\,\,\,\, \mbox{and} \,\,\,\,\nabla\theta'-\vctr{A}'=\nabla\theta-\vctr{A}.\ee Then it is easy to verify that 
\be\psi'=\psi\e{ie(\theta'-\theta)}\,\,\,\, \mbox{and} \,\,\,\,\vctr{A}'=\vctr{A}+\nabla(\theta'-\theta).\ee In other words, ($\theta'-\theta$) defines a gauge transformation from $(\psi,\vctr{A})$ to $(\psi',\vctr{A}')$.
In particular, the holonomies can be recovered from the covariant derivatives of the phase:
\be 
\oint\!\mc{D}\theta \cdot \dr{\vctr{x}}=\oint\!\nabla \theta\cdot \dr{\vctr{x}}+ \oint\!\vctr{A} \cdot \dr{\vctr{x}}=\oint\!\vctr{A} \cdot \dr{\vctr{x}},
\ee
since the integral of a gradient around a closed loop is zero.

The alert reader will have noticed something rather striking about this representation. Both $\rho$ and $\mc{D}\theta$ are \emph{local} features of the theory: their values at a point $\vctr{x}$ depend only on $\psi$ and $\vctr{A}$. The \vctr{A}-field alone may admit of no description which is both separable and gauge-invariant, but the $\psi$ and $\vctr{A}$ fields jointly admit of both.

Indeed, we can rewrite the Sch\"{o}dinger equation in a local and gauge-invariant way in terms of these quantites; since the method of doing so is instructive for later, I spell it out here. Firstly, let us make a choice of gauge: the \emph{unitary gauge}, in which $\psi$ is always real. (This may seem unfamiliar: gauge conditions are usually specified via  a constraint on \vctr{A} rather than $\psi$. But mathematically a gauge condition is just a condition which picks a unique element out of each equivalence class of gauge-equivalent fields, and --- again on the assumption that $\psi\neq 0$ --- the unitary gauge does that just fine. I return to its conceptual significance later.)

In the unitary gauge we can write $\psi=\rho$; the Schr\"{o}dinger equation becomes
\be 
\frac{1}{2m}\left(\nabla^2 \rho - \vctr{A}\cdot \vctr{A}\rho - 2i \vctr{A}\cdot \nabla \rho -i (\nabla \cdot \vctr{A}) \rho\right)= i \dot{\rho}.
\ee
Separating real and imaginary parts, we get
\be 
(\nabla^2 - \vctr{A}\cdot \vctr{A})\rho=0;
\ee
\be 
2 \vctr{A}\cdot \nabla \rho + (\nabla \cdot \vctr{A})\rho=2m \dot{\rho}.
\ee
Combined with the condition that the magnetic field strength $\vctr{B}$ vanishes, 
\be\nabla\times \vctr{A}=0,\ee
this is a complete and deterministic set of equations for $\rho$ and $\vctr{A}$ in the unitary gauge.

(If you are wondering how the Schr\"{o}dinger equation, which is supposed to determine the evolution of the \emph{particle}, has given rise to a joint equation for the particle probability density and the vector potential, recall that in the unitary gauge, phase information about the particle is carried by $\vctr{A}$. If this makes you start to worry that we don't have a clean separation any more between matter and magnetic degrees of freedom, hold that thought!)

To get a gauge-invariant set of equations, we just note that in the unitary gauge, $\nabla \theta=0$ and so $\mc{D}\theta=\vctr{A}$. So in this gauge, we can replace $\vctr{A}$ with $\mc{D}\theta$ to get
\be 
(\nabla^2 - (\mc{D}\theta)^2)\rho=0;\,\, 2 \mc{D}\theta \cdot \nabla \rho + (\nabla \mc{D} \theta)\rho=2 m \dot{\rho}; \,\,\nabla \times \mc{D} \theta =0.
\ee 
But this equation, being expressed entirely in terms of gauge-invariant quantities, does not depend on the unitary gauge. We have obtained a set of local, deterministic, gauge-invariant differential equations for the A-B effect.

(Isn't this just the One True Gauge strategy applied with respect to the unitary gauge? In a sense, yes: but the crucial point is that the unitary gauge, unlike any gauge determinable within vacuum electromagnetism, has the feature that  $\vctr{A}(\vctr{x})$ (and $\psi(\vctr{x})$) depend only on gauge-invariant features of an arbitrarily small neighborhood of $\vctr{x}$. So we really can regard the unitary gauge simply as a representation --- a coordinatisation, if you like --- of an underlying separable gauge-invariant reality.\footnote{I am grateful to Owen Maroney for pressing this point.})

All this ought to suggest that the apparent nonlocal-action/ gauge-dependence/ non-separability trilemma of the A-B effect is just an artefact of our failure to consider $\psi$ as well as $\vctr{A}$. Indeed, I think this suggestion is correct. Before exploring the suggestion further, though, it will be helpful to get clear just how that trilemma arises and how the introduction of matter blocks it.

\section{Origins of non-separability}\label{origins}

Recall the definition of non-separability: the state of a region of space $X \cup Y$  is nonseparable if specification of all properties of regions $X$ and $Y$ separately does not completely specify the properties of $X \cup Y$. In the case of electromagnetic gauge theory under  the assumption that all physical properties are gauge-invariant, the properties of a region are supposed to be in some way represented by gauge-invariant features of the fields, with two regions having the same physical properties iff the fields on those regions are gauge-equivalent.

We can now express the presence or absence of non-separability mathematically: fields $\psi,\vctr{A}$ defined on $X\cup Y$ give rise to non-separability iff there exist other fields $\psi',\vctr{A}'$ defined on $X \cup Y$ such that
\begin{enumerate}
\item[(i)]$\psi'|_X,\vctr{A}'|_X$ (the restrictions of $\psi'$ and $\vctr{A}'$ to $X$) are gauge-equivalent to $\psi|_X,\vctr{A}|_X$; 
\item[(ii)] likewise $\psi'|_Y,\vctr{A}'|_Y$ and $\psi|_Y,\vctr{A}|_Y$ are gauge-equivalent; but 
\item[(iii)]$\psi',\vctr{A}'$ and $\psi,\vctr{A}$ are \emph{not} gauge-equivalent.
\end{enumerate}
For any possible state of $X \cup Y$ must be represented by some pair of fields on $X\cup Y$, and non-separability is the possibility of two such non-gauge-equivalent pairs $\psi,\vctr{A}$ and $\psi',\vctr{A}'$ whose restrictions to $X$ and to $Y$ are gauge-equivalent.

Suppose (i) and (ii) are the case. Then there exist functions $\Lambda_X$, $\Lambda_Y$ on $X$ and $Y$ respectively such that
\begin{enumerate}
\item On $X$, $\psi'=\e{i e\Lambda_X}\psi$ and $\vctr{A}'=\vctr{A}+\nabla \Lambda_X$.
\item On $Y$, $\psi'=\e{i e\Lambda_Y}\psi$ and $\vctr{A}'=\vctr{A}+\nabla \Lambda_Y$.
\end{enumerate}
It follows that on the intersection region $X \cap Y$, 
\begin{enumerate}
\item $\e{i e(\Lambda_X-\Lambda_Y)}\psi = \psi$;
\item $\nabla (\Lambda_X-\Lambda_Y)=0$.
\end{enumerate}
So $\Lambda_X-\Lambda_Y$ is a real function on $X \cap Y$ which (1) is equal to zero except where $\psi=0$ and (2) has vanishing gradient everywhere. These are strict conditions. The first can be satisfied by $\Lambda_X-\Lambda_Y\neq 2n\pi/e$ only in regions where $\psi=0$. The second entails that if $x$ and $y$ are points in $X \cap Y$ connected by a path lying within $X \cap Y$, then $(\Lambda_X-\Lambda_Y)(x)=(\Lambda_X-\Lambda_Y)(y)$. Jointly, then, the conditions can be satisfied by a function with non-vanishing gradient only if $X \cap Y$ is path-disconnected (if there are regions of $X \cap Y$ that cannot be joined by any path lying within $X \cap Y$) and if $\psi$ is zero on at least one of the connected components.

If these conditions are not satisfied, then $\Lambda_X$ and $\Lambda_Y$ agree (up to a removable $2n\pi/e$ term) on the intersection of $X$ and $Y$. We can then define a single function $\Lambda$ consistently by declaring it equal to $\Lambda_X$ on $X$ and to $\Lambda_Y$ on $Y$; this function generates a gauge transformation between $\psi,\vctr{A}$ and $\psi',\vctr{A}'$, so that (iii) is not satisfied. 

Conversely, if they \emph{are} satisfied then we can choose arbitrary functions $\Lambda_X,\Lambda_Y$ which are constant on each connected component of $X \cap Y$ but which are not equal to each other on at least one such component. The fields obtained by applying a gauge transformation generated by $\Lambda_X$ to the restriction of $\psi,\vctr{A}$ to $X$, and likewise for $Y$, agree on $X \cap Y$ and so can be consistently combined into a pair of fields on $X\cup Y$, but they are not gauge-equivalent.

So we have found a necessary and sufficient condition for non-separability in gauge theory: it can occur with respect to regions $X$, $Y$ when their intersection is disconnected and when the matter field vanishes on at least one connected component. (In fact, the result generalises straightforwardly to more general gauge theories: what is required there is not per se that $\psi$ vanishes on a connected component but that there is some element of the gauge group $g$ such that $g \psi=\psi$ on that region. This generally requires $\psi$ to remain strictly confined to some small subspace of the internal vector space.)

The first of these conditions is purely topological. A necessary (though not sufficient) condition for it to occur is that $X \cup Y$ is not simply connected;\footnote{Proof sketch: suppose $X\cup Y$ is simply connected and let $f$ be any smooth function which is constant on each connected component of $X \cap Y$. Then for arbitrary $a$, $b$, there is a well-defined vector field $v$ on $X \cup Y$ such that $ v|_X=a \nabla f$ and $v|_Y=b \nabla f$. For arbitrary $p,q\in X \cap Y$, let $\gamma_X$ and $\gamma_Y$ be paths in $X$ and $Y$ respectively from $p$ to $q$. Then the integral of $v$ along the loop from $x$ to $y$ along $\gamma_X$ and back along $\gamma_Y$ is $(a-b)(f(q)-f(p))$. But since $\nabla \times v=0$, by Stokes' theorem this integral must vanish. So $f(p)=f(q)$, \iec any function constant on the connected components is constant.} note that this is satisfied by the region outside the solenoid in the A-B effect, and recall that we have seen that non-separability occurs in the loop ontology only where non-simply-connected regions are considered.

The second condition, however --- the vanishing of $\psi$ on an open set --- is implausibly, indeed unphysically, stringent. Notice that there is no `give' in the condition at all: even if $|\psi|=10^{-1000}$, there is no prospect of non-separability. (The local facts about $X$ and $Y$ separately that determine the joint properties of $X\cup Y$ might be extremely difficult to ascertain, but that is a limit of practice, not principle.) In one-particle quantum mechanics, it is a theorem\footnote{The result is proved under rather general conditions by Hegerfeldt~\citeyear{hegerfeldta,hegerfeldtb}; see also the discussion in \citeN{halvorsonclifton}. To see intuitively why it is correct, just notice that to confine a particle exactly to a finite region requires it to have arbitrarily high-momentum Fourier components, corresponding to arbitrarily high momenta, and so to components of the wavefunction that will spread out at arbitrarily high speeds.} that $\psi$ is never exactly zero on an open set in space\emph{time}, so that the condition can hold, if at all, only for an instant. And in quantum field theory the most perspicuous way (in this context) to think of the system is as a superposition of different field configurations, in which the weight given to the configuration where $\psi$ is \emph{exactly} zero will itself be exactly zero. (I consider the quantum-field-theoretic case more carefully in section~\ref{vacpol}). I conclude that we can set aside this case. Once set aside, there is no obstacle to a fully local, but fully gauge-invariant, understanding of the theory.

\section{The interlinking of $A$ and $\psi$}\label{interlinking}

I have shown formally that the gauge-invariant features of $\psi$ and $\vctr{A}$ can generically be jointly represented in a fully local (\iec, non-separable) way. But it is still reasonable to ask what those gauge-invariant features are actually supposed to represent: that is, what kind of ontology is compatible with the theory?

It is tempting to think that the question can be innocently rephrased as: what kind of ontologies for the electromagnetic field, and for the matter field, are compatible with the theory? Tempting, but mistaken --- and this is one of the main points of the paper. For since the gauge transformation thoroughly mixes the two together, there is simply no justification --- as long as we wish our ontology to depend only on gauge-independent features of the theory --- in regarding the two mathematically-defined fields as representing two \emph{separate} but interacting entities, rather than as (somewhat redundantly) representing aspects of a \emph{single} entity.

To press the point, let us consider again the question of a choice of gauge. Most gauge choices encountered in  electromagnetism impose a constraint on the \vctr{A}-field, and leave the $\psi$-field unconstrained: thus the Coulomb gauge, $\nabla \cdot \vctr{A}=0$, for instance, or the London gauge $\vctr{A}_z=0$ (each with an appropriate boundary condition) each place one constraint on $\vctr{A}$ per point of space. Hence the temptation to see the \vctr{A}-field, with its apparent three degrees of freedom per space point, as really having two once gauge redundancy is allowed for, and likewise to see the $\psi$ field as genuinely having two degrees of freedom per space point.

But this is pure convention. Consider again the unitary gauge, in which we require that the phase of $\psi$ vanishes (\iec, that $\psi$ is real). In this gauge, $\psi$ has only one degree of freedom, but there is no residual gauge invariance of \vctr{A} --- each of its three apparent degrees of freedom are physical. So do we have one degree of freedom for matter and three for electromagnetism, or two for each? The question is only meaningful if we persist in supposing that two distinct entities are present.

To be sure, from the perspective of quantum field theory there is no conventionality about the \emph{particles} that are associated with the fields: whatever gauge we choose, we will discover a particle spectrum consisting of a massless vector boson (two degrees of freedom) and a charged scalar boson (one degrees of freedom, but with both matter and antimatter versions\footnote{For more on the curious way in which complex classical degrees of freedom give rise to antimatter, see \citeN{wallaceantimatter} and \citeN{BakerHalvorson2009}.}). But the particle spectrum of a theory represents the expansion of the theory's Hamiltonian in normal modes around a (possibly local) minimum of energy, and is by its nature holistic: the particle spectrum of the theory is a dynamical and not a metaphysical matter, and should not be thought to require the existence of metaphysically distinct matter and electromagnetic fields.

Indeed, it need not always be the case that a complex-scalar-field-plus-vector-potential field theory even has that particular particle spectrum. If the gauge symmetry is spontaneously broken (that is, if the minimum-energy configuration has a non-zero expected value of $|\psi|$) then the particle spectrum consists instead of a massive vector field and a real scalar field (indeed, this is one of the main applications of the unitary gauge). In popularisations of the Higgs mechanism, this phenomenon is sometimes described as the electromagnetic field ``eating'' one of the degrees of freedom of the scalar field and thus gaining mass, a metaphor that has been sharply criticised by  \citeN{earmancurie} (see also \citeN{struyvegauge}). But once we realise that the electromagnetic and scalar fields cannot be thought of as separate entities, there need be no residual surprise that the normal-mode expansion of the physical system that they jointly describe is best analysed in different ways in different regimes.

But how are we to think about this ``jointly described'' entity? We know that it can be characterised entirely by the magnitude of $\psi$ (a scalar field) and by its covariant derivative (a vector field, or more precisely a one-form field). It is important to remember that these are conceptually and mathematically very different entities. A scalar field, mathematically, is just an assignment of a real number to every point of space, and can easily enough be thought of as ascribing properties to \emph{points} of space. A one-form field is not so simple and cannot be so represented: to speak loosely, it is more like an assignation of properties to infinitesimally small diffences between points of space. Or put another way, if a vector is thought of loosely as an infinitesimal arrow from one space point to a neighboring one, a one-form field assigns a real number to each such infinitesimal arrow. A one-form is then something more like 
a set of relations between (infinitesimally close) points of space. 

That suggests that there are indeed two components of the ontology of the system: a collection of properties of points of space, and a collection of relations between infinitesimally close points of space. In certain circumstances (mathematically, when the holonomy vanishes) integrating the infinitesimal relations from $x$ to $y$ along a given path gives a result which is in fact independent of the path; in these situations we can consistently define a relation between those finitely-separated points and call it the \emph{phase difference}, and then the system can be represented by a complex field with no remaining redundancy save for a single choice of phase. Conversely, the holonomy --- the integral of the infinitesimal relations around a closed loop --- provides a measure of the extent to which this representation of the systems is blocked, and the holonomy in turn is \emph{mostly} determined by the integral of the relations around infinitesimal closed loops --- the curvature.

The extent to which this somewhat loose talk of `infinitesimal relations' can be made more precise lies beyond the scope of this paper; it is perhaps worth remembering, though, that in any case the empirical success of (classical or quantum) electrodynamics provides no licence whatever to regard the theory as a reliable description of the physical world on \emph{arbitrarily short} lengthscales, so that thinking about the relations between extremely but finitely close points of space may actually be a more reliable way of approaching the theory's ontology than appeal to vector bundles or to actual infinitesimals.\footnote{For more consideration of the metaphysics of vector fields, see Butterfield~\citeyear{Butterfield2006a,Butterfield2006} and references therein.}

\section{Generalisation to non-Abelian gauge theories}\label{nonabelian}

(This section lies outside the general flow of the paper; it presumes familiarity with non-Abelian gauge theories and can be skipped by readers lacking such familiarity.)

So far I have taken $\psi$ to be a complex scalar field and $\vctr{A}$ to be a $U(1)$ connection, i.e (in the most straightforward mathematical representation) a $u(1)$-valued one-form. The more general case would take $\psi$ to lie in some representation of a compact Lie group \mc{G} and $\vctr{A}$ to be a one-form taking values in the Lie algebra $\vctr{g}$ of \mc{G}. In the gauge transformation (\ref{gt}), $\Lambda$ is now a $\vctr{g}$-valued field, and its exponential $\e{i\Lambda}$ a \mc{G}-valued field; the charge $e$ is most conveniently incorporated into the definition of \vctr{A} in this formalism.

Under this reinterpretation, the analysis of the first part of section \ref{origins} still goes through: in particular, the conditions 
\begin{enumerate}
\item $\e{i e(\Lambda_X-\Lambda_Y)}\psi = \psi$;
\item $\nabla (\Lambda_X-\Lambda_Y)=0$
\end{enumerate}
remain jointly necessary and sufficient for non-separability to arise. (2) has the same consequences as in the Abelian case: it requires that the connection is flat in the connected components of the overlap region $X \cap Y$. To understand (1), consider in general the conditions under which $R(g)v=0$ for $R$ a (faithful) vector representation of \mc{G}, $g\in\mc{G}$, and $v$ an element of the vector space on which $R$ is defined. Clearly two possible conditions are $g=\mathrm{id}$ and $v=0$; are there others?

In some cases, yes: for instance, suppose that \mc{G}=$SU(N)$ or $U(N)$ for $N>1$ and that $\psi$ is a scalar field (with the standard representation of \mc{G}). Then there will be a (N-1)-dimensional set of field values at each point invariant under any given nonidentity element of $\mc{G}$. Once again, though, this set of values is of measure zero in the full set of field and connection values.

Furthermore, in realistic non-Abelian theories even this case does not arise. Non-Abelian fields in the Standard Model are spinors, not scalars, and can be thought of as $m$-tuples of vectors acted on by \mc{G}. In this case, only the special case where all the $m$ vectors are aligned (and in particular the still-more special case $\psi=0$) will permit any distinct $\Lambda_X,\Lambda_Y$ such that (1) is satisfied.

I conclude that insofar as the $\psi=0$ special case can be set aside for Abelian fields it may likewise be set aside in the non-Abelian case: non-Abelian gauge theories, like Abelian gauge theories, can be understood locally and separably.

There is, however, one important disanalogy. The quantities $|\mc{D}\theta|$ and $|\psi|$, while perfectly well defined in the non-Abelian case at least when $\psi$ is a scalar field, do not suffice jointly to determine the gauge-invariant properties: the phase angle $\theta$ takes values in $\mc{G}$ and not $U(1)$, and is not adequately constrained by the absolute value of $\mc{D}\theta$. And while specification of $\mc{D}\theta$ itself (along with $|\psi|$) \emph{does} so suffice, the latter is not gauge-invariant. In general, I know of no comparably simple set of local gauge-invariant quantities in the non-Abelian case that can serve as a gauge-invariant representation. 

Furthermore, this is not simply an issue of \emph{non}-Abelian gauge theories. Even in the Abelian case, if we consider a spinor-valued matter field, or (equivalently) a pair of scalar matter fields $\psi_1$ and $\psi_2$ coupled to the same $U(1)$ connection, again the straightforward specification of $|\mc{D}\theta|$ and $|\psi|$ is not available. We could of course specify, say, $|\psi_1|$, $\psi_2$ and $|\mc{D}\theta_1|$, but any such stipulation is somewhat arbitrary.

On reflection this should not be surprising. A gauge field is a complex mathematical object defined with relatively little available background structure; in general (as we have learned from the Hole Argument) such objects tend to lack simple geometric descriptions and must be described, at the cost of some arbitrariness, by a particular choice of coordinatisation: in the case of gauge fields, by a particular choice of gauge. But --- as I was at pains to stress in section \ref{review} --- \emph{this} is not the paradoxical feature of the A-B effect. What is apparently paradoxical is that a gauge-invariant understanding of the ontology of electromagnetism seems to violate separability or locality. Appropriate consideration of the role of $\psi$ removes this problem in non-Abelian gauge theories as surely as in electromagnetism. 

As for what \emph{positive} account of the ontology of non-Abelian gauge theories can be given: so far as I can see, the account of section \ref{interlinking} goes over to this case too. I postpone development of this to future work, however.

\section{Vestiges of nonlocality?}\label{vestiges}

In the remainder of the paper I wish to address two possible concerns with the picture of electromagnetic ontology I have sketched. The first is whether --- for all that it is formally separable (as demonstrated in section~\ref{origins} --- it is not still in some sense non-local. For consider again the regions $X$ and $Y$ around the solenoid in figure 1. It seems that absolutely nothing about the gauge-invariant state of region $X$ alone is in any way dependent on the strength of the magnetic flux in the solenoid. After all, that region is characterised by a flat connection! And likewise, absolutely nothing about region $Y$'s state seems to depend on the flux. And yet the joint state of $X \cup Y$ does depend (through its holonomy) on the flux. So is there not still an important sense in which the physics is non-separable?

As I will show in section \ref{vacpol}, it is not strictly correct to say that the states of $X$ and $Y$ separately are not dependent on the flux. But put that aside for now; it \emph{is} correct, at any rate, in the usual semiclassical accounts of the A-B effect. It remains true that in physically realistic situations the $\psi$ field will not be exactly zero on an open set in either $X$ or $Y$, and this will be the key to resolving our puzzle.

For the connection on $X$ defines a parallel-transport rule across $X$, and since $X$ is simply connected, we can speak without ambiguity of the phase difference between any two points in $X$. In particular, there will be some phase difference $\theta_X$. between $\psi$ at a point $\alpha$ on one side of the region in the overlap with $Y$, and at another point $\beta$ in the overlap on the other side, and this phase difference will be a physical --- and in principle physically \emph{measurable} --- property of $X$ alone. In terms of our previous gauge-invariant description, $\theta_X$ is the integral of the (gauge-invariant) covariant derivative $\mc{D}\psi$ of $\psi$ along a path from $\alpha$ to $\beta$ within $X$. 

Similarly, there will be a phase difference $\theta_Y$ between those two points within $Y$, equal to the integral of $\mc{D}\psi$ along a path joining them within $Y$, and this will be a physical property of $Y$ alone that again can in principle be measured. But the difference $\theta_X$ and $\theta_Y$ is just the integral of $\mc{D}\psi$ around the solenoid, which in turn --- since the integral of $\nabla \psi$ around any closed loop is zero --- is just the electric charge times the holonomy around the solenoid. So $\theta_X-\theta_Y$ is proportional to the magnetic flux, and so that flux is jointly determined by properties of $X$ and $Y$. Even though there is no straightforward dependence of the properties of each region separately on the flux, there is a co-dependence of their properties jointly on the flux.

It's worth pausing to see how this differs from the case where $\psi$ is neglected. There the integral of $\vctr{A}$ around the closed loop from $\alpha$ to $\beta$ and back again is determined by the flux, but the open integrals of $\vctr{A}$ from $\alpha$ to $\beta$ through either $X$ or $Y$ are not. But the difference is that those open integrals, unlike the open integrals of $\mc{D}\psi$, are \emph{gauge-dependent}, and so  (if we accept gauge equivalence as physical equivalence) do not determine any physical property of $X$ or $Y$. When we neglect $\psi$ we get a non-separability of \emph{ontology}; when we allow for it, we get only what might be called a non-separability of \emph{counterfactual dependence}. But really even this overstates the case: it is not that the dependency is really non-separable but just that nothing \emph{general} can be said about the separate counterfactual dependencies of $\theta_X$ and $\theta_Y$. For any given physical situation things can perfectly well be different (I will return to this shortly). 

It is instructive to compare the ``non-locality'' here with that of the so-called ``gravitational A-B effect'' \cite{Dowker1967}. Here we suppose that space contains a very long (in idealisation, infinite) thin cylinder of very high curvature, such that the region around the cylinder --- though spatially flat --- has the property that a circle of radius $R$ centred on the cylinder has circumference not $2\pi R$ but $(2 \pi- \Delta) R$, with the value of $\Delta$ depending on the curvature within the cylinder. (In general relativity, open cosmic strings --- so far unobserved --- produce such geometries.) If restricted to two dimensions, such a space has the geometry of a cone: curved only at the tip, but such that the angle through which it is necessary to move to go completely around the tip is $2\pi - \Delta$ radians, not $2\pi$ radians.

We can again divide the region around the cylinder into overlapping sub-regions $X$ and $Y$. Each region is simply connected and geometrically flat, so that nothing internal to it distinguishes it from flat space. But there are facts about each region which jointly determine everything about the combined region $X \cup Y$. Specifically, consider the field of radial lines going directly outwards from the cylinder (such a field can be defined using only local resources). Parallel-transporting a vector through $X$ from one to the other regions of overlap with $Y$ will result in that vector undergoing a rotation $\theta_X$ relative to the field of radial lines; likewise, parallel-transporting a vector through $Y$ will result in a relative rotation $\theta_Y$. Neither $\theta_X$ nor $\theta_Y$ will itself depend on the degree of curvature within the cylinder, but their difference $\theta_X-\theta_Y$ is equal to $\Delta$ and thus depends upon the curvature. Like the A-B effect as described here --- but unlike the A-B effect when $\psi$ is neglected --- the degree of geometric nontriviality of the region around the cylinder is determined by physically measurable features of the separate parts of that region.

\section{What is the particle interacting with?}\label{interactingwithwhat}\label{vacpol}

I have defended a unified account of the ontology represented by the $\vctr{A}$ and $\psi$ fields. But this seems to conflict with a fairly central feature of the A-B effect as empirically demonstrated. Namely, there seems to be some entity which is generated by the magnetic field and in virtue of which the space around the solenoid has a disposition to cause phase shifts in charged particles that are diffracted around the solenoid. And there is some other entity --- the charged particle --- which is indeed so diffracted and which causes the disposition to be manifest. This certainly sounds an awful lot like an interaction between two systems, and the mathematics of the A-B effect makes those systems look an awful lot as though they are represented (perhaps redundantly) by the \vctr{A} and $\psi$ fields respectively.

I think this problem really brings to mind the awkwardness of the standard description of the A-B effect, which I have been glossing over thus far. As standardly described, the effect concerns the evolution of a quantum-mechanical particle --- represented by a wave function --- under the influence of the classical electromagnetic potential; it is thus a semi-classical phenomenon which we would expect to be awkward at best to describe in a fully consistent manner. In particular, it is widely recognised that there is in general no satisfactory account of the interaction of quantum and classical systems that involves back-reaction of the quantum on the classical system (note that in standard treatments of the A-B effect the $\vctr{A}$-field is taken as background and as being unaffected by the $\psi$-field). It is then unsurprising that an analysis at this level builds in a dichotomy between the two systems that may not correctly represent the true goings on.  

In my discussions of the ontology so far I indulged in the fiction that both $\psi$ and $\vctr{A}$ are fully classical entities, but this analysis --- though consistent --- does not describe the A-B effect as actually observed. To really understand the A-B effect from an ontological point of view we need to find a consistent account of electromagnetic  fields interacting with \emph{quantum-mechanical} matter, and this account, of course, requires quantum field theory (QFT).\footnote{For an attempt to develop a consistent account within the framework of nonrelativistic quantum mechanics, see \citeN{vaidmanAB}; Vaidman also concludes that the A-B effect properly understood is local, but it is not immediately clear to me how his approach connects to this paper's.}

From a QFT perspective it is relatively easy to identify the two distinct entities that are interacting in the A-B effect. Firstly,we have the lowest-energy (``ground'') field state of the region around the solenoid, treating the current within the solenoid either as a classical background condition or as a Gaussian state heavily peaked around a macroscopically large current, but in either case taking it as a background assumption.  This state describes the physical goings-on around the solenoid in the absence of any particles. (In relativistic particle physics the ground state is often taken to be the true lowest-energy state of the system; in more general applications, such as condensed-matter physics, the ground state may instead be only the lowest-energy state given some constraint.)

Secondly we have particles, which in QFT are the normal-mode excitations of the quantum field around its ground state. These excitations can in general be formed into wave-packets, are in general precisely defined only to the extent that the field is non-interacting, and in that latter case there is an isomorphism between the one-particle Hilbert space and the space of solutions to the classical field equations (see \citeN{wallaceantimatter} for further discussion). A particle wave-packet state can be prepared distant from the solenoid region and then sent to interact with that region; there is no \emph{completely} precise separation of the two entities but to a very high degree of accuracy we can regard the solenoid region and the wave packet as separate entities which interact with one another.

However, none of this relies on any particular separation of \vctr{A} and $\psi$. The ground state of a QFT, and the excitations of that ground state, are \emph{global} features of the theory, and in particular there is no sense in which the solenoid region state is one where only the $\vctr{A}$ field, and not the $\psi$ field, is affected by the presence of the current in the solenoid. 

Indeed, we can coherently ask for the expectation value of the gauge-invariant observables $\op{\rho}$ and $\op{\mc{D}\psi}$ 
in the vicinity of the solenoid. They will in general be dependent on the flux through the solenoid even outside the solenoid itself; that is, in general the solenoid leads to vacuum polarisation in the region around it. Calculating this vacuum fluctuation is not especially trivial (see \citeN{Serebryanyi1985} and \citeN{Gornicki1990}) but we can get some handle on it by considering the quantity $\langle \op{\rho}^{-1}\op{\mc{D}\psi}\rangle$: the expected value of the covariant derivative of the phase, which we can calculate explicitly: by Ehrenfest's theorem, its integral around a closed loop surrounding the solenoid will equal the charge $e$ times the expected value of the magnetic flux through the solenoid (which is just its classical value $\Phi$) modulo $2 \pi$, with the lowest-energy case being predictable on physical grounds as corresponding to the lowest possible such value. If we adopt cylindrical polar coordinates $(r,\theta,z)$, and write $\mc{D}_\theta$ for the covariant derivative in the direction of increasing $\theta$, we have
\be 
\int \dr{\theta}r \langle \op{\rho}(r,\theta,z)^{-1}\op{\mc{D}_\theta\psi}(r,\theta,z)\rangle=[e\Phi] 
\ee
where I write $[e\Phi]$ for $e\Phi$ modulo $2\pi$.
But the problem is symmetric under rotations and translations in $z$ (under the idealisation that the solenoid is infinite) so the expression in the integral does not in fact depend on $\theta$ or $z$, and we have
\be 
\langle \op{\rho}(r)^{-1}\op{\mc{D}_\theta\psi}(r)\rangle=\frac{[e\Phi]}{2\pi r}.
\ee
So the expected value of the covariant derivative of the phase, which on symmetry grounds can be expected to vanish in free space, diverges from zero by an amount proportional to the A-B phase shift. 
This vacuum polarisation can be thought of as a quantum-mechanical version of the reference $\psi$-field used in the previous section: it is an entirely local, and in principle\footnote{To the best of my knowledge this measurement has not been carried out} measurable, feature of the space around the solenoid knowledge of which suffices to calculate the level of interference-fringe shift caused by the A-B effect.

\section{Conclusion}\label{conclusion}

I have argued that the apparent clash between separability and gauge invariance in the A-B effect is an artefact of failing to allow for the $\psi$ field as well as the $\vctr{A}$ field, and for the existence of joint gauge-invariant features of the two which do not reduce to gauge-invariant features of either considered separately. I have demonstrated that except in the (classically pathological, quantum-field-theoretically irrelevant) case where the $\psi$ field vanishes exactly on an open set, the magnitude of the $\psi$ field and the covariant derivative of its phase --- both of which are entirely local and entirely gauge-independent quantities --- suffice to fix all gauge-invariant features of the theory. I have argued that the ontological lesson to draw is not that the theory is non-separable. Instead, the lesson is that  $\vctr{A}$ and $\psi$ do not represent (however redundantly) separate entities, but rather that they carry out their representational task jointly (and redundantly). And I have shown how this is to be reconciled with the actual physics of the A-B effect, and identified particular local features of the space around the solenoid (vacuuum polarisation) which suffice to determine the magnitude of the induced phase shift.

There is something ironic about this whole story. As long as the electromagnetic field interacted with charged matter only through equations which could be written in terms of $\vctr{B}$ alone, of course an ontology involving $\vctr{B}$ alone sufficed. It was only when we coupled that field to a complex scalar field in a gauge-invariant way that the theory became sensitive to features which transcend $\vctr{B}$, but it is precisely the introduction of that complex scalar field which also keeps the theory separable while remaining gauge-invariant. In hindsight (that most unhelpfully wise of philosophical tools) attempting an analysis of the ontology of electromagnetism which required us to preserve the $\vctr{B}$-transcending features while disregarding the extra component which actually incorporated those features into a locally-stateable dynamics could have been predicted to give rise to an inaccurate understanding of the situation.

I close with a remark aimed at a different and more general area of physics. The ontology I have suggested for the electromagnetic field in the light of the A-B effect is not a dualistic ontology of $\psi$ and $\vctr{A}$, but it \emph{is} dualistic for all that, consisting both of properties of space (or spacetime) points and of (infinitesimal) relations between those points. This way of thinking about the ontology of contemporary field theory --- and in particular, the recognition that these two aspects of the ontology do not map neatly onto the two mathematical objects used to represent them --- potentially has value in thinking both about gauge theories more generally and about contemporary theories of spacetime. I hope to expand upon this observation in future work.

\section*{Acknowledgements}

This paper has benefitted from useful conversations with Harvey Brown, Richard Healey, Eleanor Knox, Owen Maroney, Simon Saunders, Chris Timpson, and a variety of undergraduates over several years of teaching, in particular Paulina Sliwa and Neil Dewar.

\end{document}